\documentclass[english,showpacs,floatfix]{revtex4}

\usepackage[dvips]{graphicx}
\usepackage{longtable}
\usepackage{amsmath,amssymb}
\usepackage{dcolumn}
\usepackage{latexsym}
\usepackage{babel}
\usepackage[latin1]{inputenc}

\begin{document}
\title{Quasinormal modes and phase Transition of black holes}
\author{Xiaoping Rao}
\author{Bin Wang}
\email{wangb@fudan.edu.cn}
\affiliation{Department of Physics, Fudan University, Shanghai
200433, People's Republic of China }

\author{Guohong Yang}
\affiliation{Department of Physics, Shanghai University, People's Republic of China }


\begin{abstract}
We have studied the scalar field as well as the fermonic field perturbations in the background of the massless BTZ black holes.
Comparing with the perturbation results in the generic nonrotating BTZ black hole background, we found that
the massless BTZ hole contains only normal modes in the perturbations. We argued that this special property reflects that the massless
BTZ black hole is a different phase from that of the generic nonrotating BTZ hole.

\end{abstract}

\maketitle

%%%%%%%%%%%%%%%%%%%%%%%%%%%%%%%%%%%%%%%%%%%%%%%%%%%%%%%%%%%%%%%%%%%%%%%%%%%%%%%
Black holes' quasinormal modes (QNM) have been an intriguing subject of discussions in the past decades \cite{1,2,3}.
The QNMs is believed as a characteristic sound of black holes, which describes the damped oscillations under perturbations in
the surrounding geometry of a black hole with frequencies and damping times of the oscillations entirely fixed
by the black hole parameters.
The QNMs of black holes has potential astrophysical interest since it could lead to the direct identification of
the black hole existence through gravitational wave
observation to be realized in the near future\cite{1,2}.
Motivated by the discovery of the AdS/CFT correspondence, the investigation of QNM in anti-de
Sitter(AdS) spacetimes became appealing in the past several years. It was argued that the QNMs
of AdS black holes have direct interpretation in term of the dual conformal field theory(CFT)\cite{3,4,5,6,7,8,9}.
Attempts of using QNMs to investigate the dS/CFT correspondence has also been given\cite{10}.
Recently QNMs in asymptotically flat spaces have acquired further attention, since the possible
connection between the classical vibrations of a black hole spacetime and various quantum aspects
was proposed by relating the real part of the QNM frequencies to the Barbero-Immirzi(BI) parameter,
a factor introduced by hand in order that loop quantum gravity reproduces correctly the
black hole entropy \cite{11}. The extension has been done in the dS background \cite{12},
however in the AdS black hole spacetime, the direct relation has not been found\cite{13}.
Further motivation of studying the QNMs has been pointed out in a very recent paper arguing that QNMs can
reflect the black hole phase transition\cite{14}. By calculating the QNMs of electromagnetic perturbations,
in \cite{14} it was claimed that they found the evidence
of the phase transition in the QNMs behavior for small topological black holes with scalar hair once disclosed
in \cite{15}.

The motivation of the present paper is to further explore the possibility of disclosing the black hole phase transition
in its QNMs behavior. We will concentrate our attention on the three dimensional spacetimes.
The mathematical simplicity in the three-dimensional cases can help us to understand the physics better.
The non-rotating $\left( {J = 0} \right)$
 BTZ black hole is described by the
line element
 \begin{equation} ds^2  =  - \left( { - M + \frac{{r^2 }}{{l^2 }}} \right)dt^2  +
\left( { - M + \frac{{r^2 }}{{l^2 }}} \right)^{ - 1} dr^2  + r^2
d\phi ^2 .\label{one}
\end{equation}
It possesses a continuous mass spectrum to the
massless AdS black holes $\left( {M = 0} \right)$
 with different topology
\begin{equation}
ds^2  =  - \frac{{r^2 }}{{l^2 }}dt^2  + \frac{{l^2 }}{{r^2 }}dr^2
+ r^2 d\phi ^2 ,\label{two}
\end{equation}
where we find a degenerate event horizon at the origin of the
coordinate $r = 0$ . Cai et al observed that when $M\rightarrow
0$, in the microcanonical ensemble some second moments
diverge\cite{16}. The divergence of the second moments means that
the fluctuation is very large which breaks down the rigorous
meaning of thermodynamical quantities. This is the characteristic
of the point of phase transition\cite{17}. The massless BTZ black
hole has zero Hawking temperature, zero entropy and vanishing heat
capacity, which is the same corresponding to that of the usual
extreme holes. For the extreme BTZ black hole, its wave dynamics
behavior has been compared to that of the nonextreme BTZ hole in
\cite{chile}. In \cite{18} it was shown that the massless BTZ
black hole persists supersymmetries, while the generic nonrotating
BTZ black hole does not, which further manifested that the
massless hole and the gereric BTZ hole are two different phases.
The massless hole is a critical point which seperates the generic
nonrotating BTZ black hole from the AdS space \cite{19}. We would
like to investigate the wave dynamics in the massless BTZ hole
background and examine whether the perturbation can help to
disclose that it is a different phase from that of the generic
nonrotating BTZ hole. To obtain a general and solid result, we
will first reexamine the scalar perturbation which was
investigated in \cite{Lee}, then we will extend our discussion to
the fermionic perturbation of the massless BTZ black hole. Besides
the study of the perturbation dynamics in the bulk, we will also
study from the CFT side. We will compare our results in the
massless BTZ background to those obtained in the non-rotating BTZ
hole \cite{6,7}.

For the massless BTZ hole, the scalar perturbation can be described
by
\begin{equation}
\left[ {\frac{1}{{\sqrt { - g} }}\partial _\mu  \left( {\sqrt { -
g} g^{\mu \nu } \partial _\nu  } \right) - \mu _{_0 }^2 }
\right]\Phi  = 0, \nonumber
\end{equation}
where $\mu_0$ is the mass of the scalar field.
Using the seperation of the wave function
$\Phi \left( {t,r,\phi } \right) = \frac{1}{{\sqrt r }}R\left( r
\right)e^{ - i\omega t} e^{im\phi }$,
the radial wave equation becomes
\begin{equation}
r^2 R^{''} \left( r \right) + 2rR^{'} \left( r \right) + l^2
\left( {\frac{{l^2 \omega ^2  - m^2 }}{{r^2 }} - \mu _{_0 }^2  -
\frac{3}{{4l^2 }}} \right)R\left( r \right) = 0. \label{twentytwo}
\end{equation}
The solution of the radial equation is given as a linear combination
\begin{equation}
R\left( r \right) = AR^{\left( 1 \right)} \left( r \right) +
BR^{\left( 2 \right)} \left( r \right), \label{twentythree}
\end{equation}\\
where the function $R^{\left( 1 \right)} \left( r \right)$
and $R^{\left( 2 \right)} \left( r \right)$ read
\begin{equation}
R^{\left( 1 \right)}  = \left( {\frac{1}{r}} \right)^{\frac{1}{2}
+ \beta } e^{ - \frac{{i\alpha }}{r}} F\left( {\frac{1}{2} + \beta
,1 + 2\beta ,\frac{{2i\alpha }}{r}} \right),
\end{equation}
\begin{equation}
R^{\left( 2 \right)}  = \left( {\frac{1}{r}} \right)^{\frac{1}{2}
- \beta } e^{ - \frac{{i\alpha }}{r}} F\left( {\frac{1}{2} - \beta
,1 - 2\beta ,\frac{{2i\alpha }}{r}} \right).
\end{equation}
$\alpha  = l\sqrt {\omega ^2 l^2  - m^2 } $,
$\beta  = \sqrt {\mu _{_0 }^2 l^2  + 1} $
and $F\left( {a,c,z} \right)$ is the confluent hypergeometric function (Kummer's
solution). $F\left( {a,c,z} \right)=1+\Sigma^{\infty}_{n=1}\frac{(\alpha)_n}{n!(\gamma)_n}z^n$ and $(\alpha)_0=1; (\alpha)_n=
\alpha(\alpha+1)\cdots(\alpha+n-1)$.

The massless BTZ hole has the same boundary condition as that of the generic BTZ hole that at infinity there is no outgoing wave due to the
infinite effective potential. If $\omega  =  \pm \frac{m}{l}$ , then $\alpha=0$ , we
obtain
\begin{equation}
R^{\left( 1 \right)}  = \left( {\frac{1}{r}} \right)^{\frac{1}{2}
+ \beta } ,
\end{equation}
\begin{equation}
R^{\left( 2 \right)}  = \left( {\frac{1}{r}} \right)^{\frac{1}{2}
- \beta } .
\end{equation}
The boundary condition requires $B$ in (\ref{twentythree}) must be zero. For general case,
the constant $B$ also requires to be zero in the spacial infinity
and the asymptotic limit of the wave function becomes
\begin{equation}
\Psi _\infty   = \frac{1}{{\sqrt r }}A\left( {\frac{1}{r}}
\right)^{\frac{1}{2} + \beta } e^{ - \frac{{i\alpha }}{r}} e^{ -
i\omega t} e^{im\phi }  = Ae^{ - i\omega t} e^{im\phi } e^{ -
\frac{{i\alpha }}{r}} \frac{1}{{r^{1 + \beta } }}.
\end{equation}
The required boundary condition is automatically satisafied, which shows that contrary to the generic black hole case where
QNMs are proportional to the quantized imaginary part of the frequency, the QNMs in the massless BTZ hole is absent.
There is only normal modes in the perturbation for the massless BTZ hole.

To be more careful, we also check whether the flux vanishes at the infinity.
The conserved radial current can be calculated as
\begin{equation}
J\left( r \right) = \Phi ^* \left( r \right)\frac{d}{{dr}}\Phi
\left( r \right) - \Phi \left( r \right)\frac{d}{{dr}}\Phi ^*
\left( r \right).
\end{equation}
Because of the boundary condition, at the infinity, the only
contribution to the current $J(r)$  comes from $R^{\left( 1
\right)} \left( r \right)$. Using the second Kummer formula
$e^{ - \frac{z}{2}} F\left( {\alpha ,2\alpha ,z} \right) = {}_0F_1
\left( {\frac{1}{2} + \alpha ;\frac{{z^2 }}{{16}}} \right)$, $R^{\left( 1 \right)} \left( r \right)$
can be written as
\begin{equation}
R^{\left( 1 \right)}  = \left( {\frac{1}{r}} \right)^{\frac{1}{2}
+ \beta } e^{ - \frac{{i\alpha }}{r}} F\left( {\frac{1}{2} + \beta
,1 + 2\beta ,\frac{{2i\alpha }}{r}} \right) = \left( {\frac{1}{r}}
\right)^{\frac{1}{2} + \beta } {}_0F_1 \left( {1 + \beta ;\; -
\frac{{\alpha ^2 }}{{4r^2 }}} \right).
\end{equation}
Thus the conserved radial current is $J(r)=0$ and the flux is given by
\begin{equation}
F = \sqrt g \frac{1}{{2i}}J\left( r \right) = 0. \label{thirtythree}
\end{equation}
This was also observed in the study of the scalar perturbation in
\cite{Lee}. The vanishing of the flux at infinity confirms the
absence of QNMs for the massless BTZ hole under the scalar
perturbation. This result is independent of the mass of the scalar
field.

To make our result general and solid, we extend our discussion to
the fermonic perturbation in the background of the massless BTZ
black hole. The Dirac equation
\begin{equation}
\gamma ^a e_a^\mu  \left( {\partial _\mu   + \Gamma _\mu  }
\right)\Psi  = \mu _s \Psi
\end{equation}
can be written in the form
\begin{equation}
\left( {\frac{{i\sigma ^2 }}{{\sqrt f }}\partial _t  + \sigma ^1
\sqrt f \partial _r  + \frac{{\sigma ^3 }}{r}\partial _\phi   +
\frac{1}{l}\sigma ^1 } \right)\Psi  = \mu _s \Psi,
\end{equation}
where $\mu_s$ is the mass of the field.
Employing the wave seperation
\begin{equation}
\Psi \left( {t,r,\phi } \right) = e^{ - i\omega t} e^{im\phi }
\left( {\begin{array}{*{20}c}
   {H\left( r \right)}  \\
   {G\left( r \right)}  \\
\end{array}} \right). \label{sixtyeight}
\end{equation}
we obtain
\begin{equation}
\left( {\frac{{i\sigma ^2 }}{{\sqrt f }}\partial _t  + \sigma ^1
\sqrt f \partial _r  + \frac{{\sigma ^3 }}{r}\partial _\phi  }
\right)e^{ - i\omega t} e^{im\phi } \left( {\begin{array}{*{20}c}
   {H\left( r \right)}  \\
   {G\left( r \right)}  \\
\end{array}} \right) = \mu _s e^{ - i\omega t} e^{im\phi } \left( {\begin{array}{*{20}c}
   {H\left( r \right)}  \\
   {G\left( r \right)}  \\
\end{array}} \right).
\end{equation}\\
Substituting the Pauli matrics, we have
\begin{equation}\left\{ {\begin{array}{*{20}c}
   {r^2 G^{'} \left( r \right) + \left( {r - i\omega l^2 } \right)G\left( r \right) = \left( {\mu _s r - im} \right)lH\left( r \right)}  \\
   {r^2 H^{'} \left( r \right) + \left( {r + i\omega l^2 } \right)H\left( r \right) = \left( {\mu _s r + im} \right)lG\left( r \right)}  \\
\end{array}} \right. \label{seventyfive}
\end{equation}
Taking $Z_ \pm  (r) = G(r) \pm H(r)$, the equation
(\ref{seventyfive}) can be rewritten as
\begin{equation}
\left\{ {\begin{array}{*{20}c}
   {r^2 Z_ + ^{'}  + (1 - \mu _s l)rZ_ +   = il(\omega l + m)Z_ -  }  \\
   {r^2 Z_ - ^{'}  + (1 + \mu _s l)rZ_ -   = il(\omega l - m)Z_ +  }  \\
\end{array}} \right.\label{seventysix}
\end{equation}
whose solution reads
\begin{equation}
\left\{ {\begin{array}{*{20}l}
   \begin{array}{l}
 Z_ +   = Cr^{ - 3/2} e^{ - \frac{{i\alpha }}{r}} \left( {\frac{{2i\alpha }}{r}} \right)^{\frac{1}{2} + \nu _ +  } F\left( {\frac{1}{2} + \nu _ +  ,1 + 2\nu _ +  ,\frac{{2i\alpha }}{r}} \right) \\
  \;\;\;\;\;\;\;    + Dr^{ - 3/2} e^{ - \frac{{i\alpha }}{r}} \left( {\frac{{2i\alpha }}{r}} \right)^{\frac{1}{2} - \nu _ +  } F\left( {\frac{1}{2} - \nu _ +  ,1 - 2\nu _ +  ,\frac{{2i\alpha }}{r}} \right) \\
 \end{array}  \\
   \begin{array}{l}
 Z_ -   = i\sqrt {\frac{{\omega l - m}}{{\omega l + m}}}  \cdot
   \left[ {Cr^{ - 3/2} e^{ - \frac{{i\alpha }}{r}} \left( {\frac{{2i\alpha }}{r}} \right)^{\frac{1}{2} + \nu _ -  } F\left( {\frac{1}{2} + \nu _ -  ,1 + 2\nu _ -  ,\frac{{2i\alpha }}{r}} \right)} \right.\;\;\;\;\; \\
     \;\;\;\;\;\;\;\;\;\;\;\;\;\;\;\;\;\;\;\;\;\;\; + \left. {Dr^{ - 3/2} e^{ - \frac{{i\alpha }}{r}} \left( {\frac{{2i\alpha }}{r}} \right)^{\frac{1}{2} - \nu _ -  } F\left( {\frac{1}{2} - \nu _ -  ,1 - 2\nu _ -  ,\frac{{2i\alpha }}{r}} \right)} \right]\\
 \end{array}  \\
\end{array}} \right.\label{seventyseven}
\end{equation}\\
where $\nu _ \pm   = \frac{1}{2} \pm \mu _s l$.

We see in (\ref{seventyseven}) that the vanishing boundary condition at infinity can be
automatically satisfied provided that $\mu _s l <3/2$. The boundary condition
puts a limit on the mass of the fermionic field, which has also been observed in \cite{7}.
In th Dirac modes, as done in \cite{7}, we can also impose vanishing flux at infinity.
The radial current is defined by
\begin{equation}
J\left( r \right) = \Psi ^* \frac{d}{{dr}}\Psi  - \Psi
\frac{d}{{dr}}\Psi ^* .
\end{equation}
Using the wave function (\ref{sixtyeight}), the current becomes
\begin{equation}
J\left( r \right) = H^* \frac{d}{{dr}}H - H\frac{d}{{dr}}H^*  +
G^* \frac{d}{{dr}}G - G\frac{d}{{dr}}G^* .
\end{equation}\\
Employing $Z_ \pm  (r) = G(r) \pm H(r)$ , we have
\begin{equation}
\begin{array}{l}
 \;\;\;\;\;\;\;\;\;\;\;\;\;\;\;\;\;\;\;\;\;\;\;\;\;\;\;\;\;\;\;\;\;\;\;\;\;\;\;\;\;\;\;\;Z_ + ^* \frac{d}{{dr}}Z_ +   - Z_ +  \frac{d}{{dr}}Z_ + ^*  + Z_ - ^* \frac{d}{{dr}}Z_ -   - Z_ -  \frac{d}{{dr}}Z_ -^*\\
 = \left( {G^*  + H^* } \right)\frac{d}{{dr}}\left( {G + H} \right) - \left( {G + H} \right)\frac{d}{{dr}}\left( {G^*  + H^* }\right)
 + \left( {G^*  - H^* } \right)\frac{d}{{dr}}\left( {G - H} \right) - \frac{d}{{dr}}\left( {G - H} \right)\left( {G^*  - H^* }
      \right)\\
   = 2\left( {G^* \frac{d}{{dr}}G - G\frac{d}{{dr}}G^*  + H^* \frac{d}{{dr}}H - H\frac{d}{{dr}}H^* } \right) \\
 \end{array}
\end{equation}
thus
\begin{equation}
\begin{array}{l}
 J\left( r \right) = H^* \frac{d}{{dr}}H - H\frac{d}{{dr}}H^*  + G^* \frac{d}{{dr}}G - G\frac{d}{{dr}}G^*
    = \frac{1}{2}\left( {Z_ + ^* \frac{d}{{dr}}Z_ +   - Z_ +  \frac{d}{{dr}}Z_ + ^*  + Z_ - ^* \frac{d}{{dr}}Z_ -   - Z_ -  \frac{d}{{dr}}Z_ - ^* } \right). \\
 \end{array}
\end{equation}
From equation (\ref{seventysix}), we get
\begin{equation}
Z_ -   = \frac{{r^2 \frac{d}{dr}Z_ +   + (1 - \mu _s l)rZ_ +
}}{{il(\omega l + m)}}.
\end{equation}
It is easy to show that
\begin{equation}
 Z_ - ^* \frac{d}{{dr}}Z_ -   - Z_ -  \frac{d}{{dr}}Z_ - ^*
= \frac{{\omega l-m}}{{\omega l + m}}\left( {Z_ + ^* \frac{d}{{dr}}Z_ +   - Z_ +  \frac{d}{{dr}}Z_ + ^* } \right).
\end{equation}\\
Using
\begin{equation}
\begin{array}{l}
 Z_ +   = Cr^{ - 3/2} e^{ - \frac{{i\alpha }}{r}} \left( {\frac{{2i\alpha }}{r}} \right)^{\frac{1}{2} + \nu _ +  } F\left( {\frac{1}{2} + \nu _ +  ,1 + 2\nu _ +  ,\frac{{2i\alpha }}{r}} \right)
    + Dr^{ - 3/2} e^{ - \frac{{i\alpha }}{r}} \left( {\frac{{2i\alpha }}{r}} \right)^{\frac{1}{2} - \nu _ +  } F\left( {\frac{1}{2} - \nu _ +  ,1 - 2\nu _ +  ,\frac{{2i\alpha }}{r}} \right) \\
 \;\;\;\; \;\;= Cr^{ - 3/2} \left( {\frac{{2i\alpha }}{r}} \right)^{\frac{1}{2} + \nu _ +  } {}_0F_1 \left( {1 + \nu _ +  ; - \frac{{\alpha ^2 }}{{4r^2 }}} \right) + Dr^{ - 3/2} \left( {\frac{{2i\alpha }}{r}} \right)^{\frac{1}{2} - \nu _ +  } {}_0F_1 \left( {1 - \nu _ +  ; - \frac{{\alpha ^2 }}{{4r^2 }}} \right) \\
 \end{array}
\end{equation}
which is a purely real function, thus we have
\begin{equation}
J\left( r \right)\; = \frac{{\omega l}}{{\omega l + m}}\left( {Z_
+ ^* \frac{d}{{dr}}Z_ +   - Z_ +  \frac{d}{{dr}}Z_ + ^* } \right)
= 0.
\end{equation}
So the flux is $F = \sqrt g \frac{1}{{2i}}J\left( r \right) = 0$. This shows that even if we do not use the
usual boundary condition for the BTZ hole which puts limit on the Dirac field mass, the weaker condition that flux vanishing
at infinity can also be satisfied automatically. This result is very different from that of the generic nonrotating BTZ hole which
requires the nonzero imaginary part of the frequency in the perturbation to meet the boundary conditions.

According to the AdS/CFT correspondence, the QNMs of the AdS black
hole can also be got from CFT. For the usual BTZ black hole, the
CFT study on the QNMs has been done in \cite{7}. For the massless
BTZ hole, we can carry out the CFT study on the QNMs as follows.
After doing the coordinate transformation $w_{\pm}=u_{\pm}=\psi\pm
t/l, y=l/r$, we can rewrite the metric of the massless BTZ hole in
the form \cite{Esko}
\begin{equation}
ds^2  = \frac{{l^2 }}{{y^2 }}\left[ {dy^2  + dw_ +
dw_ - } \right].\label{poincare0}
\end{equation}

A massive scalar field $\Psi$ with a mass $\mu_0$ in this
spacetime behaves as $\Psi\rightarrow y^{2h_-}\Psi_0 (w_+,w_-)$
near the boundary $y\rightarrow 0$. The boundary field
$\Psi_0(w_+,w_-)$ is of mass dimension $2h_-$ and acts as a source
to an operator $O(w_+,w_-)$ of mass dimension $2h_+$ in boundary
theory, where $h_{\pm}=(1\pm \sqrt{1+\mu_0^2})/2$.

The bulk field $\Psi(y,(w_+,w_-))$ can be obtained from the
boundary field $\Psi_0(w_+,w_-)$ through
\begin{equation}
\Phi \left( {y,(w_ {+}  ,w_ {-})  } \right) = \int {dw_{+} ^{'} }
dw_ {-} ^{'} G \left( {y,w_ {+}  ,w_ {-}  ;w_ {+} ^{'} ,w_ {-}
^{'} } \right)\Phi _0 \left( {w_ {+} ^{'} ,w_ {-} ^{'} }
\right),\label{bulkboundary}
\end{equation}
where the bulk-boundary Green's function is
\begin{equation}
G \left( {y,w_ {+}  ,w_{ -}  ;w_ {+} ^{'} ,w_{ -} ^{'} } \right) =
c\left( {\frac{y}{{y^2 + \Delta w_ +  \Delta w_ -  }}}
\right)^{2h_ +  },
\end{equation}
with $\Delta w_ \pm   = w_ \pm   - w_ {\pm} ^{'} $ .

In the massless BTZ hole, we have the bulk-boundary Green's
function
\begin{equation}
G_{M = 0} \left( {r,u_ +  ,u_ -  ;u_ {+} ^{'} ,u_{ -} ^{'} }
\right) = c\left( {\frac{r}{{1 + r^2 \Delta u_ +  \Delta u_ -  }}}
\right)^{2h_ +  },
\end{equation}
with $\Delta u_ \pm   = u_ \pm   - u_{_ \pm  }^{'} $. The relation
between the bulk field and the boundary field now reads
\begin{equation}
\Phi \left( {r,u_ +  ,u_ -  } \right) = \int {du_ {+ }^{'} } du_
{-} ^{'} G_{M = 0} \left( {r,u_ + ,u_ -  ;u_ {+} ^{'} ,u_ {-} ^{'
}} \right)\Phi _0 \left( {u_{ +} ^{'} ,u_{ -} ^{'} }
\right).\label{mbulkboundary}
\end{equation}

Evaluating the surface integral \cite{E3}
\begin{equation}
I\left( \Phi  \right) = \mathop {\lim }\limits_{r_s  \to \infty }
\int_{T_s } {du_ +  du_ -  \sqrt h \Phi \left( {\hat e_r  \cdot
\nabla } \right)\Phi } ,\label{surfaceintegral}
\end{equation}
where  $T_s$ is the surface $r=r_s$ , $h$  its induced metric of
the massless BTZ hole and $\vec e_r$  is the unit vector normal to
the surface, one gets in $r\rightarrow \infty$
\begin{equation}
I\left( \Phi  \right) \sim \int {du_ +  du_ -  du_ {+} ^{'} } du_
{- }^{'} \Phi _0 \left( {u_ +  ,u_ -  } \right)\left(
{\frac{1}{{\Delta u_ +  \Delta u_ -  }}} \right)^{2h_ +  } \Phi _0
\left( {u_ {+} ^{'} ,u_ {-} ^{'} } \right).
\end{equation}
The two point correlator can be read as
\begin{equation}
\left\langle {O\left( {u_ +  ,u_ -  } \right)O\left( {u_ {+} ^{'}
,u_ {-} ^{'} } \right)} \right\rangle  \sim \left(
{\frac{1}{{\Delta u_ + \Delta u_ -  }}} \right)^{2h_ +  }.
\end{equation}

For the perturbations, we have
\begin{equation}\begin{array}{l}
 \int {du_ +  du_ -  du_ {+} ^{'} } du_{-} ^{'} \Phi _0 \left( {u_ +  ,u_ -  } \right)\frac{1}{{\left( {\Delta u_ +  \Delta u_ -  } \right)^{2h_ +  } }}\Phi _0 \left( {u_{+} ^{'} ,u_{-} ^{'} } \right) \\
 \;\; = \int {du_ +  du_ -  du_{ +} ^{'} } du_ {-} ^{'} \frac{{e^{i\left( {p_ +  u_ +   + p_ -  u_ -  } \right) + i\left( {p_ {+} ^{'} u_{ +  }^{'}  + p_ {-} ^{'} u_{ -  }^{'} } \right)} }}{{\left( {u_ +   - u_ {+ }^{'} } \right)^{2h_ +  } \left( {u_ -   - u_ {- }^{'} } \right)^{2h_ +  } }} \\
 \;\; \sim \delta \left( {p_ +   - p_{+} ^{'} } \right)\delta \left( {p_-   - p_{- }^{' }} \right)\left( {\frac{1}{{\Gamma \left( {2h_ +  } \right)}}}
 \right)^2,
 \end{array}\end{equation}
where $p_ \pm   = \frac{1}{2}\left( {m \mp \omega } \right)$. In
getting the last step above, we have employed
\begin{equation}
\frac{1}{{\Gamma \left( z \right)}} = \frac{1}{{2\pi }}\int_{
- \infty }^\infty  {e^{a + iu} \left( {a + iu} \right)} ^{ - z}
du.
\end{equation}
There is no pole in the above expression. In the field theory, the
poles in the momentum representation of the retarded correlation
function corresponds to the QNMs\cite{7}. Without the pole means
that the QNMs is absent in the massless BTZ hole, which is in
consistent with the perturbation wave dynamics studied in the
bulk. The CFT result should also hold in the fermonic
perturbation.

In summary in this work we have shown that different from that of
the non-rotating BTZ black hole\cite{6,7}, the frequency of the
perturbations in the massless BTZ black hole has only the real
part. There is only normal modes in the perturbations. The
vanishing boundary conditions in the spacetime at infinity can
automatically be satisfied for the massless BTZ solution, which
implies that the QNMs in the massless BTZ black hole is absent.
This is different from that of the generic nonrotating BTZ black
hole where the boundary condition can only be satisfied for
appropriate quantized fenquency with the nonzero imaginary part.
We argue that the special fields' perturbation results in the
massless BTZ black hole reflects that it is a different phase from
that of the non-rotating BTZ black holes. Our result could serve
as a support to \cite{16}.

%%%%%%%%%%%%%%%%%%%%%%%%%%%%%%%%%%%%%%%%%%%%%%%%%%%%%%%%%%%%%%%%%%%%%%%%%%%%%
\begin{acknowledgments}
%%%%%%%%%%%%%%%%%%%%%%%%%%%%%%%%%%%%%%%%%%%%%%%%%%%%%%%%%%%%%%%%%%%%%%%%%%%%%
This work was partially supported by  NNSF of China, Ministry of
Education of China and Shanghai Science and Technology Commission.
\end{acknowledgments}

\end{document}